**Title:** Seismic Hazard and Public Safety

**Author**: Lalliana Mualchin (1)

(1) International Seismic Safety Organization
Viale San Francesco,12,
I-64031 Arsita (TE), Italy.

**Abstract**
International Seismic Safety Organization (ISSO) has been formed to promote public safety by being prepared for the largest potential events which can happen at any time, rather than for certain probable events which have been exceeded in several recent earthquakes with disastrous consequences. The position of ISSO is available now in English, Italian, Russian, Hebrew, Spanish, and Hindi at http://www.issoquake.org. That position has been misrepresented elsewhere and this short note is to counter such inaccurate viewpoints.

**Introduction**
Warner Marzocchi (2013) has discussed extensively the position of the International Seismic Safety Organization (ISSO) and its implications in a Forum article but we regret that it is a clear misrepresentation that can mislead the readers. The concern of ISSO has been primarily the repeated failures of probabilistic seismic hazard estimates in earthquake-prone regions where recent events exceeded such estimates again and again (e.g., Wyss and others, 2012), and thus exposing the public to such hazard and unacceptable risk, from under-developed countries (e.g., Haiti) to most advanced countries (e.g., Italy & Japan). Engineers have poor opinions about probabilistic approach (e.g., Gulcan, 2013), suggested stop funding such studies (Klugel, 2014), and scientists found that the results or maps are erroneous (Kossobokov and Nekrasova, 2010; 2012).

It is at this juncture that ISSO has been formed for promoting a solution as presented in its position.

**Discussions**
The ISSO does not favor using recurrent time as a factor in specifying earthquake magnitudes based on assumed time-independent mean rate or temporal Poisson distribution, for earthquake source strength for scaling its hazards as applied in the probabilistic seismic hazard analysis (PSHA); and strongly recommends using the ***largest*** potential earthquake size based on the spatial extent and style of the fault that is anticipated for earthquake process within the time scale of modern tectonics (e.g., Holocene or late Quaternary). This is the Maximum Credible Earthquake (MCE) magnitude as applied in the deterministic seismic hazard analysis (DSHA) and neo-DSHA (NDSHA) which can happen at any time (see, Mualchin, 2010 for the history of its successful use in California; and Peresan and Panza, 2012 for Italy).

Using that time scale as the limiting recurrent time for MCE by Marzocchi (2013) is manipulating the premise that any seismic hazard analysis must be based on accepting the process as somehow repeatable. The repeatable process does not always follow PSHA's recurrent model, for example, even mega-earthquakes cluster in time. In millions of years, earthquake sources, including subduction zones may appear and disappear at various locations many times, making any local estimate of earthquake potential "time-dependent" at this scale of time. Note that seismic hazard analysis is done within the time scale as stated above that effectively excludes earthquake sources not currently existing.

There are important reasons for using the respective preferred magnitude inputs by PSHA and DSHA that Marzocchi (2013) has not discussed for an objective evaluation of the approaches. PSHA was an attempt to equitably consider various faults according to its seismic activities characterized by recurrence and/or slip rates derived from earthquakes and faults studies. DSHA is interested in considering the largest potential earthquakes on seismic sources with no regard to recurrences or slip rates. These purposes define the inputs and how the results have been presented. Unfortunately, DSHA has not been promoted in the literature and misinterpreted by PSHA proponents for a number of years, including Marzocchi's (2013) viewpoints. Probabilistic magnitudes are simply not determined as scenario deterministic ones for any sized events; recurrence determines the former but not the latter, even if both are of the same size. Without elaborations, recent findings point out that the concept of earthquake recurrence (viz., the seismic cycle) can be misleading and even unrealistic (Bizzarri and Crupi, 2013).

It is high time that the public be given DSHA viewpoints as they need to know for making informed decisions. It should be noted that adequate estimates of ground motions must consider the tensor nature of seismic wave generation and propagation process and include required information for risk assessment using appropriate statistical models as incorporated in NDSHA (e.g., Panza et al., 2013).

There is an undefined uncertainty in estimating earthquake recurrence due both to the limitations in present knowledge and the fact that geologic processes, such as faster or slower fault slip movements do not operate on at a steady rate even though probabilities may be assumed for the recurrence. *However, it is an established fact that low or high probability earthquakes can occur at any time, in terms of seismic history and reality but not in modelling.* We cannot reject the hypothesis of another M9.0 earthquake occurrence within the next few years in the area of the 11 March 2011 Tōhoku mega-earthquake, even though the chance may be very small.

This is precisely the reason why ISSO strongly recommends the explicit use of MCE magnitudes for critical cases when the consequences of failure are intolerable. This is the only sensible approach for establishing seismic hazard for critical structures, such as nuclear power plants, dams, hospitals and schools for public safety. By its very concept and purpose, PSHA generally downgrades hazard from MCEs due to its alleged long recurrences for design purposes. The superiority of PSHA for public safety has not been demonstrated but numerous problems have been found that can be simply ignored by regulators from their safe position. Obviously, much more issues are involved than can be given in this limited space for our response that ISSO can provide in the future. Neglecting this principle can cause again catastrophic consequences, as documented in recent earthquakes in 2008 in China; 2009 in Italy; 2010 in Haiti; 2010 and 2011 in New Zealand; and 2011 in Japan. The ongoing costly and painfully prolonged disaster of the Fukushima nuclear power-plant is a spectacular example of ignoring the largest potential earthquake and the associated tsunamis for the design, construction, and management of a nuclear power plant.

The criticisms of ISSO based on arguments using recurrence, return period, and such have no basis because, even if a more reliable evaluation of recurrence for NDSHA estimates is possible (Peresan et al., 2013), those are not considered in our position. There is an implicit assumption in Marzocchi's (2013) arguments that earthquake occurrence is Poissonian so that observed data represent the situation "for all time" to obtain the probability of occurrence for any time interval, no matter how long. For example, available earthquake data may be only for 50 years but the calculated hazard may be projected to any number of years, say 2,000 years. There is no basis for such results, no matter how sophisticated the analysis may be. Observations are to be interpreted within the time period of observations and not to be supplanted by analysis beyond that time as apparent in PSHA.

Seismic hazard may not be mitigated but the associated risks can be mitigated by reducing vulnerability. Therefore, seismic risk mitigation should focus on optimizing the cost for reducing vulnerability, rather than manipulating the seismic hazard by using recurrence, and choosing arbitrary return periods, etc. That is, choosing longer return periods correspond to higher hazards and shorter periods to lower hazards. For example, the 1000-year or the 100-year return periods ground motion hazards can happen at any time. There are no objective criteria for choosing return periods to be easily manipulated even though the calculations may be real. Hazard is directly related to the size of the event, which does not follow Poissonian distribution in the real world. Moreover, PSHA misrepresented seismic hazard for use in building codes because its poor performance versus seismic reality has not been informed to engineers and code officials who created the codes.

ISSO strongly recommends defining seismic hazards by using the ***largest*** potential or MCE magnitudes based on the spatial extend and style of faults, and not based on recurrence, return periods, etc. It should be noted that seismic hazard assessments based on MCE have been successfully used in California for design and risk mitigation since the early 1970's and are the only approach to assure the necessary public safety. The estimated hazards near the sources for the 1994 M6.7 Northridge and 1992 M7.3 Landers earthquakes in southern California are on the same order as manifested by these events according to the deterministic map but much lower per USGS probabilistic map (Mualchin, 1996). The failure of probabilistic map was considering these sources as relatively inactive with long recurrence and the success of deterministic was considering the MCEs for these faults. It became a surprise event for probabilistic but anticipated by the deterministic approach. Even though events smaller than

MCEs may be more frequent, preparing for the MCEs will automatically consider them by virtue of its most impactful size. It does not make good sense to prepare for smaller ones just because they may be expected more frequently statistically and not prepared for the MCEs for public safety, as they would be most destructive.

**Summary**
ISSO has been formed to promote public safety by being prepared for the largest potential hazardous event that can be scientifically estimated as it would be most impactful and can happen at any time. Hazardous events based on recurrent times or probabilities are not to be used for public safety matters inspite of Marzocchi's (2013) faulty arguments.

**Acknowledgments**
Without mentioning their names, members and friends of the International Seismic Safety Organization (ISSO) have generously contributed in the preparation of this note. Thanks to the publisher (arXiv.org) for posting this heartfelt note on their very dynamic site.